\documentclass{aa}    
\usepackage{txfonts}
\usepackage{longtable}  
\usepackage{natbib}
\usepackage{graphicx}

\bibpunct[, ]{(}{)}{,}{a}{}{,}

\usepackage{color}

\newcommand{\sneden}{CS~22892-052} 
\newcommand{\cayrel}{CS~31082-001} 
\newcommand{\cs}{CS~29497-004}                     
\newcommand{\LyudmilasStar}{HE~2327-5642}        
 
\newcommand{\tefft}{$T_{\mbox{\scriptsize eff}}$}  

\newcommand{\eps}[1]{\log\varepsilon_{\rm #1}}
\newcommand{\kms}{km\,s$^{-1}$}

\newcommand{\iso}[2]{\mbox{$^{#1}{\rm #2}$}}
\newcommand{\kH}{$S_{\!\!\rm H}$}    

\begin{document}

\title{The Hamburg/ESO R-process Enhanced Star survey (HERES) \thanks{Based on
    observations collected at the European Southern Observatory, Paranal,
    Chile (Proposal numbers 170.D-0010, and 280.D-5011).}}
\subtitle{IX. Constraining pure r-process Ba/Eu abundance ratio \\ from observations of r-II stars}

\author{
  L. Mashonkina\inst{1,2} \and
  N. Christlieb\inst{3}
}

\offprints{L. Mashonkina; \email{lima@inasan.ru}}
\institute{
     Universit\"ats-Sternwarte M\"unchen, Scheinerstr. 1, D-81679 M\"unchen, 
     Germany \\ \email{lyuda@usm.lmu.de}
\and Institute of Astronomy, Russian Academy of Sciences, Pyatnitskaya st. 48, 
     RU-119017 Moscow, Russia \\ \email{lima@inasan.ru}
\and Zentrum f\"ur Astronomie der Universit\"at Heidelberg, Landessternwarte,
     K\"onigstuhl 12, D-69117 Heidelberg, Germany \\
     \email{N.Christlieb@lsw.uni-heidelberg.de}
}

\date{Received  / Accepted }

\abstract
{The oldest stars born before the onset of the main $s$-process are expected to reveal a pure $r$-process Ba/Eu abundance ratio.}
{We revised barium and europium abundances of selected very metal-poor (VMP) and strongly $r$-process enhanced (r-II) stars to evaluate an empirical $r$-process Ba/Eu ratio.}
{Our calculations were based on 
  non-local thermodynamic equilibrium (NLTE) line formation for 
  \ion{Ba}{ii} and \ion{Eu}{ii} in the classical 1D MARCS model atmospheres. Homogeneous stellar abundances were determined from the \ion{Ba}{ii} subordinate and resonance lines by applying a common Ba isotope mixture. We used high-quality VLT/UVES spectra and observational material from the literature.}
{For most investigated stars, NLTE leads to a lower Ba, but a higher Eu abundance. The resulting elemental ratio of the NLTE abundances amounts, on average, log(Ba/Eu) = 0.78$\pm$0.06. This is a new constraint to pure $r$-process production of Ba and Eu. The obtained Ba/Eu abundance ratio of the r-II stars supports the corresponding Solar System $r$-process ratio as predicted by recent Galactic chemical evolution calculations of Bisterzo, Travaglio, Gallino, Wiescher, and K{\"a}ppeler. We present the NLTE abundance corrections for lines of \ion{Ba}{ii} and \ion{Eu}{ii} in the grid of VMP model atmospheres.}
{}


\titlerunning{Constraining pure $r$-process Ba/Eu ratio from observations of the r-II stars}
\authorrunning{Mashonkina \& Christlieb}

\maketitle
%
\section{Introduction}\label{Sect:intro}

Since the classical paper of \citet{B2FH} one traditionally believes that heavy elements beyond the iron group are produced in the neutron-capture nuclear reactions, which can proceed as the slow ($s$) or rapid ($r$) process. The distribution of the $s$-process abundances of the Solar System (SS) matter has been recognized as arising from a non-unique site. 
    The isotopic distribution in the atomic mass range A = 90-208 is accounted for by the main $s$-process. It occurs in intermediate-mass stars of 1-4~$M_\odot$ during the asymptotic giant branch (AGB) phase, and it is rather well understood theoretically \citep[see, for example][and references therein]{1999ARA&A..37..239B}. 
    The weak $s$-process runs in helium burning core phase of massive stars ($M > 10 M_\odot$) and it contributes to the nuclei up to A = 90 \citep{1989RPPh...52..945K}. 
 The $r$-process takes place in an extremely n-rich environment. Astrophysical sites for the $r$-process are still debated, although they are likely associated with explosions of massive stars, with  $M > 8 M_\odot$. We refer to the pioneering review of \citet{1978SSRv...21..639H} and also \citet{2004PhT....57j..47C} for further discussion on the $r$-process. 

In the Solar System matter, different isotopes were produced in differing proportions from the $s$- and $r$-process. For isotopes with A $> 90$, the $s$-abundances are evaluated from the Galactic chemical evolution (GCE) calculations. 
The difference between solar total and $s$-abundance is referred to as $r$-residual. The Solar system $r$-process (SSr) pattern is widely employed in stellar abundance comparisons \citep[see][and references therein]{Sneden2008}. However, different $s$-process calculations result in different SSr, in particular, for the chemical species with dominant contribution of the $s$-process to their solar abundances. This paper concerns with barium that together with Sr are the best observed neutron-capture elements in very metal-poor (VMP) and extremely metal-poor (EMP) stars. For example, the resonance lines of \ion{Ba}{ii} were measured in the [Fe/H]\footnote{In the classical notation, where [X/H] = $\log(N_{\rm
    X}/N_{\rm H})_{star} - \log(N_{\rm X}/N_{\rm H})_{Sun}$.} $\simeq -4.0$ stars in our Galaxy \citep{Francois2007} and also the classical dwarf spheroidal galaxies \citep{Tafelmeyer2010}. One of the most cited SSr is based on calculations of \citet[][stellar model, hereafter, A99]{Arlandini1999} who used stellar AGB models of 1.5 and 3~$M_\odot$ with half solar metallicity and predicted that 81~\%\ of the solar barium are of $s$-process origin (Table\,\ref{Tab:ba_isotope}). Very similar result was obtained by \citet[][hereafter, T99]{Travaglio1999} by the integration of $s$-abundances from different generations of AGB stars, i.e., considering the whole range of Galactic metallicities. \citet[][hereafter, B11]{Bisterzo2011} updated calculations of \citet{Arlandini1999} by accounting for the recent n-capture cross-sections, and they inferred a higher $s$-process contribution to the solar barium of 89~\%. Slightly lower solar $s$-abundance of Ba was obtained by \citet[][hereafter, B14]{2014arXiv1403.1764B} in their recent GCE calculations that considered the contributions from different generations of AGB stars of various mass. From one hand side, differences of 5 to 10~\%\ between different predictions are comparable with the quoted uncertainties in $s$-abundances (Table\,\ref{Tab:ba_isotope}). From other hand side, a small change in the $s$-abundance leads to significant change in the solar $r$-abundance of Ba and this has an important consequence for the Ba/Eu ratio of $r$-abundances, (Ba/Eu)$_r$. In contrast to Ba, solar europium is mostly composed of $r$-nuclei (Table\,\ref{Tab:ba_isotope}). The change in n-capture cross-sections shifts log(Ba/Eu)$_r$ from 0.93\footnote{In this study, the decomposition of the $s$- and $r$-process contributions is based on the meteoritic abundances of \citet{Lodders2009}.} (A99) down to 0.75 (B11). Updating the GCE calculations results in lower log(Ba/Eu)$_r$ = 0.87 (B14) compared with log(Ba/Eu)$_r$ = 0.96 of T99.
It is worth noting, the SSr data cover the full range of predictions from the $r$-process models. In the classical waiting-point (WP) approximation, \citet{Kratz2007} inferred log(Ba/Eu)$_r \simeq$ 1, and log(Ba/Eu)$_r \simeq$ 0.8 was obtained in large-scale parameterized dynamical network calculations of \citet{Farouqi2010} in the context of an adiabatically expanding high-entropy wind (HEW), as expected to occur in core-collapse SNe.
%

\begin{figure}  
  \resizebox{88mm}{!}{\includegraphics{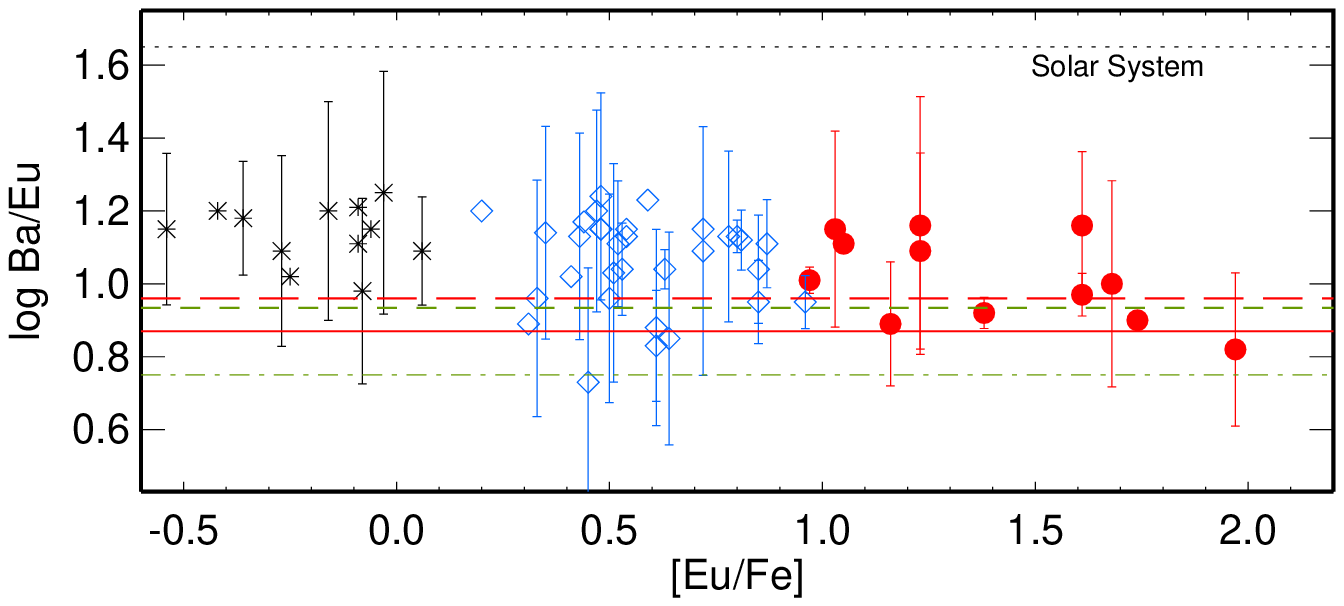}}
  \caption{\label{Fig:halo} The Ba/Eu abundance ratios of the r-II (filled circles), r-I (open rombs), and
    Eu-poor (asterisks) stars (for the sources of the data, see text). The error bars were computed as $\sigma_{\rm Ba/Eu} = \sqrt{\sigma_{\rm Ba}^2 + \sigma_{\rm Eu}^2}$, where the abundance errors available. 
 The continuous and long-dashed lines indicate the SSr ratios, as predicted by GCE calculations of \citet{2014arXiv1403.1764B} and \citet{Travaglio1999}, respectively, while the short-dashed and dash-dotted lines correspond to the SSr of \citet{Arlandini1999} and \citet{Bisterzo2011}, respectively. The dotted line corresponds to the Solar System ratio \citep{Lodders2009}.}
\end{figure}

In this paper, we investigate whether the observed stellar abundances of Ba and Eu can constrain a pure $r$-process Ba/Eu ratio.
Since the $s$- and $r$-process are associated with stars of different mass, their contribution to heavy element production varied with time. The Ba/Eu ratio is particularly sensitive to whether the $s$- or $r$-process dominated the nucleosynthesis. 
Old stars born before the onset of the main $s$-process should reveal more Eu relative to Ba compared with the SS matter. The existence of VMP stars enriched in the $r$-process element Eu was observationally established more than 30 years ago in a pioneering paper of \citet{Spite1978}, and the statistics was significantly improved in later studies \citep[for a review, see][]{Sneden2008}.
Figure\,\ref{Fig:halo} shows the Ba/Eu ratios of a preselected sample of metal-poor (MP, [Fe/H] $< -1.5$) stars that reveal enhancement of Eu relative to Ba, with [Ba/Eu] $< -0.4$. The stars are separable into three groups, depending on the observed Eu abundance. \citet{HERESI} classified the stars, which exhibit large enhancements of Eu relative to Fe, with [Eu/Fe] $> 1$, as r-II stars. The Ba and Eu abundances of the r-II stars together with the sources of data are given in Table\,\ref{Tab:stars}. The stars of r-I type have lower Eu enhancement, with [Eu/Fe] = 0.3 to 1. The data for 32 r-I stars were taken from \citet{Cowan2002,Honda2004,HERESII,Ivans2006,Francois2007,Mashonkina2007,Lai2008}, and \citet{HE1219}. Here, we also deal with the 12 stars that reveal a deficiency of Eu relative to Fe, with $\mathrm{[Eu/Fe]} \le 0$ \citep{Honda2004,Honda2006,Honda2007,HERESII,Francois2007}. They are referred to as Eu-poor stars. 
The different groups have consistent within the error bars Ba/Eu ratios, independent of the observed Eu/Fe ratio, with the mean log(Ba/Eu) = 1.03$\pm$0.12, 1.08$\pm$0.13, and 1.14$\pm$0.08 for the r-II, r-I, and Eu-poor stars, respectively. Hereafter, the statistical error is the dispersion in the single star abundance ratios about the mean: $\sigma = \sqrt{\Sigma(\overline{x}-x_i)^2/(n-1)}$. 
The observed stellar abundance ratios are more than 1$\sigma$ higher compared with the Solar System $r$-process ratio log(Ba/Eu)$_r$ = 0.87 based on recent updated $s$-process calculations of B14, although they are consistent within the error bars with the SSr of T99.

\begin{figure} 
  \centering
  \resizebox{88mm}{!}{\includegraphics{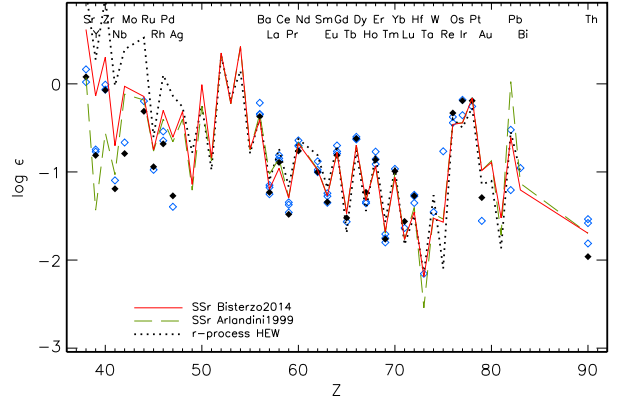}}
  \caption{\label{Fig:AbundancePattern} The heavy-element abundance patterns of 
    {\sneden} (filled rombs) and the benchmark r-II stars
    CS\,31082-001, HE\,1523-0901, and HE\,1219-0312 (open triangles). The element abundances have 
    been scaled to match Eu--Tm in {\sneden}. The dashed and continuous curves indicate the SSr abundance patterns calculated using the $s$-process predictions of 
    A99 and B14, respectively. The dotted curve corresponds to the HEW $r$-process model of \citet{Farouqi2010}. }
\end{figure}

To make clear the situation with stellar abundances of Ba and Eu, we concentrate on the r-II stars that are best candidates for learning about the details of the $r$-process and its site. Their heavy element abundances are dominated by the influence of a single, or at most very few nucleosynthesis events. The first such object, {\sneden}, with [Fe/H] = $-3.1$ and [Eu/Fe] = 1.63, was discovered by \citet{Sneden1994}. At present, 12 r-II stars are known, and the fraction of r-II stars at [Fe/H] $< -2.5$ is estimated at the level of 5\,\%\ (Paper~II). Detailed abundance analysis of {\sneden} \citep{sneden1996} and another benchmark r-II star {\cayrel} \citep{hill2002} established the match of the stellar and solar $r$-process pattern in the Ba-Hf range suggesting that the $r$-process is universal. It produced its elements with the same proportions during the Galactic history. This conclusion was of fundamental importance for better understanding the nature of the $r$-process. Figure\,\ref{Fig:AbundancePattern} displays the heavy-element abundance patterns of four stars revealing the largest Eu enhancement, with [Eu/Fe] $\ge 1.5$. 
 The data were taken from \citet[][{\sneden}]{Sneden2003}, \citet[][{\cayrel}]{2013A&A...550A.122S}, \citet[][HE\,1219-0312]{HE1219}, and \citet[][HE\,1523-091]{Sneden2008}. All these stars have very similar chemical abundance patterns in the Sr-Hf (probably Pt?) range suggesting a common origin of these elements in the classical $r$-process. Only single measurements or upper limits are available for the heavier elements up to Pb, not allowing any firm conclusion about their origin. Two of the four r-II stars have high abundances of Th, and they are referred to as actinide-boost stars. 
Figure\,\ref{Fig:AbundancePattern} displays also the SSr patterns from predictions of B14 and A99. Two sets of the solar $r$-abundances are consistent except the elements with significant contribution of the $s$-process to their solar abundances. This concerns, in particular, the light trans-Fe elements. For example, for Sr and Y, the $s$-process contribution exceeds 90\,\%. In such a case, the calculation of the $r$-residuals involves the subtraction of a large number from another large number, so that any small variation in one of them leads to a dramatic change in the difference. The uncertainty in the solar $r$-residuals does not allow to draw firm conclusions about any relation between the light trans-Fe elements in r-II stars and the solar $r$-process.
     For elements beyond Ba, the difference between SSr(B14) and SSr(A99) is notable for Ba, La, Ce, Ta, and Pb. The only measurement of stellar Ta \citep[][{\cayrel}]{2013A&A...550A.122S} favors the solar $r$-residual of B14.
Due to the weakness of the lines of the lead in the optical wavelength region, Pb abundances have been measured only in very few metal-poor stars \citep[for a recent review, see][]{Roederer2009}. This paper focuses on stellar Ba and Ba/Eu abundance ratios.

Our first concern is the line formation treatment. For cool stars, most abundance analyses are made under the assumption of local thermodynamic equilibrium (LTE), and Fig.\,\ref{Fig:halo} shows the LTE abundance ratios. In MP atmospheres, the departures from LTE can be significant due to a low number of electrons donated by metals, which results in low collision rates, and also due to low ultra-violet (UV) opacity, which results in high photoionization rates. Therefore, a non-local thermodynamic equilibrium (NLTE) line-formation modeling has to be undertaken. Each line of \ion{Ba}{ii} and \ion{Eu}{ii} consists of isotopic and hyper-fine splitting (HFS) components, and derived element abundances depend on which isotope mixture was applied in the calculations. Different $r$-process models predict different fractional abundances of the Ba isotopes, and different stellar Ba abundance analyses are based on different data on the Ba isotope mixture. We wish to evaluate systematic shifts in derived stellar Ba/Eu abundance ratios due to using different Ba isotope mixtures.
 
 \begin{table} 
 \caption{\label{Tab:ba_isotope} Solar System Ba and Eu isotope abundance fractions (\%) and $s$-process contributions (\%).}
 \centering
 \begin{tabular}{rrllcll}
\hline\hline \noalign{\smallskip} 
  & Total & \multicolumn{5}{c}{$s$-process}  \\
\cline{3-7}
\noalign{\smallskip}
 & & \multicolumn{2}{c}{1.5, 3~$M_\odot$ AGB models} & & \multicolumn{2}{c}{GCE calculations} \\
\noalign{\smallskip}
\cline{3-4}
\cline{6-7}
\noalign{\smallskip}
              & L2009& A99  & B11  & & T99 & ~~B14 \\
\noalign{\smallskip}
\hline 
\noalign{\smallskip}
\iso{134}{Ba} & 2.4  &  100 &  100 & & ~~94 & ~~100 \\
\iso{135}{Ba} & 6.6  & ~~30 & ~~26 & & ~~22 & ~~~~28 \\
\iso{136}{Ba} & 7.9  &  100 & 100  & & ~~97 & ~~100 \\
\iso{137}{Ba} & 11.2 & ~~67 & ~~66 & & ~~58 & ~~~~63 \\
\iso{138}{Ba} & 71.7 & ~~94 & ~~86 & & ~~84 & ~~~~92 \\
\multicolumn{2}{l}{Total Ba} & 81$\pm6.7$ & 88.7$\pm5.3$ & & ~~80 & 85.2$\pm6.7$ \\
\noalign{\smallskip} 
\hline 
\noalign{\smallskip} 
\iso{151}{Eu} & 47.8 & ~~~6 & ~~~6 & & ~~~6 & ~~~~~6 \\
\iso{153}{Eu} & 52.2 & ~~~5 & ~~~6 & & ~~~5 & ~~~~~6 \\
\multicolumn{2}{l}{Total Eu} & ~~~6$\pm7$ & ~~~6$\pm0.3$ & & ~~~6 & ~~~6$\pm0.4$  \\
\noalign{\smallskip}\hline \noalign{\smallskip}
\multicolumn{7}{l}{L2009 = \citet{Lodders2009}, } \\
\multicolumn{7}{l}{A99 = \citet{Arlandini1999}, B11 = \citet{Bisterzo2011},} \\
\multicolumn{7}{l}{T99 = \citet{Travaglio1999}, B14 = \citet{2014arXiv1403.1764B}.} \\
\end{tabular}
\end{table}

This paper is structured as follows. In Sect.\,\ref{Sect:nlte}, we describe NLTE calculations for \ion{Ba}{ii} and \ion{Eu}{ii} in VMP atmospheres and evaluate the HFS effects on Ba abundance determinations. Abundances of Ba and Eu of the selected r-II stars are revised in Sect.\,\ref{Sect:stars}. Section\,\ref{Sect:Conclusions} summarizes the obtained results. 

\section{Method of calculations}\label{Sect:nlte}

\subsection{NLTE modelling of \ion{Ba}{ii} and \ion{Eu}{ii}}

In the NLTE calculations for \ion{Ba}{ii} and \ion{Eu}{ii}, we used model atoms treated in our earlier studies \citep[][updated]{Mashonkina1999,mash_eu}. The coupled radiative transfer and statistical equilibrium (SE) equations were solved with a revised version of the DETAIL program
\citep{detail}, based on the accelerated lambda iteration method
described by \citet{rh91,rh92}. An update was presented by
\citet{mash_fe}. The obtained level populations were then used to calculate spectral line profiles with the code SIU \citep{Reetz}. We used classical plane-parallel (1D) models from the MARCS grid \citep{Gustafssonetal:2008}\footnote{\tt http://marcs.astro.uu.se}.

\begin{figure}
\flushleft 
\hspace{-5mm}
\resizebox{88mm}{!}{\includegraphics{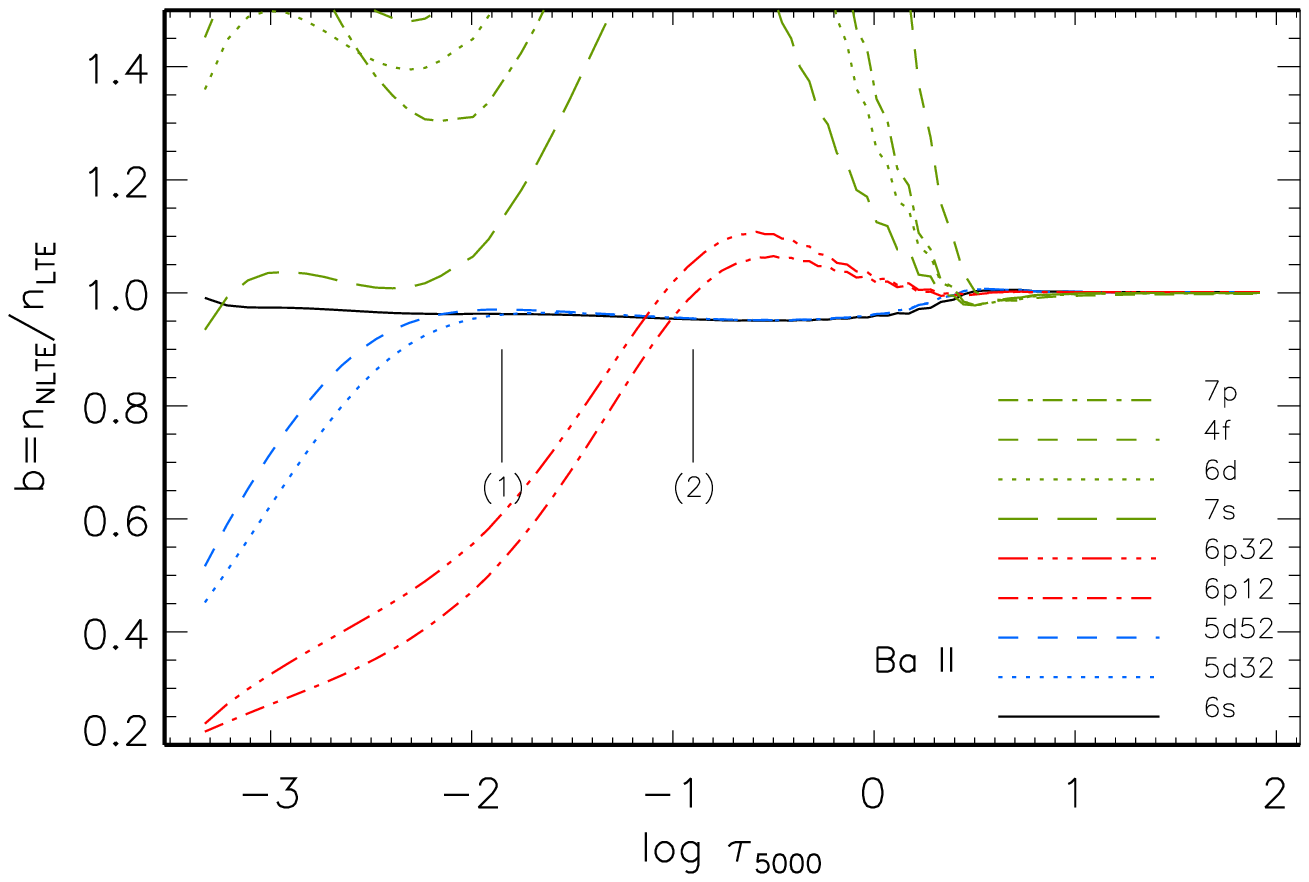}}

\hspace{-5mm}
\resizebox{88mm}{!}{\includegraphics{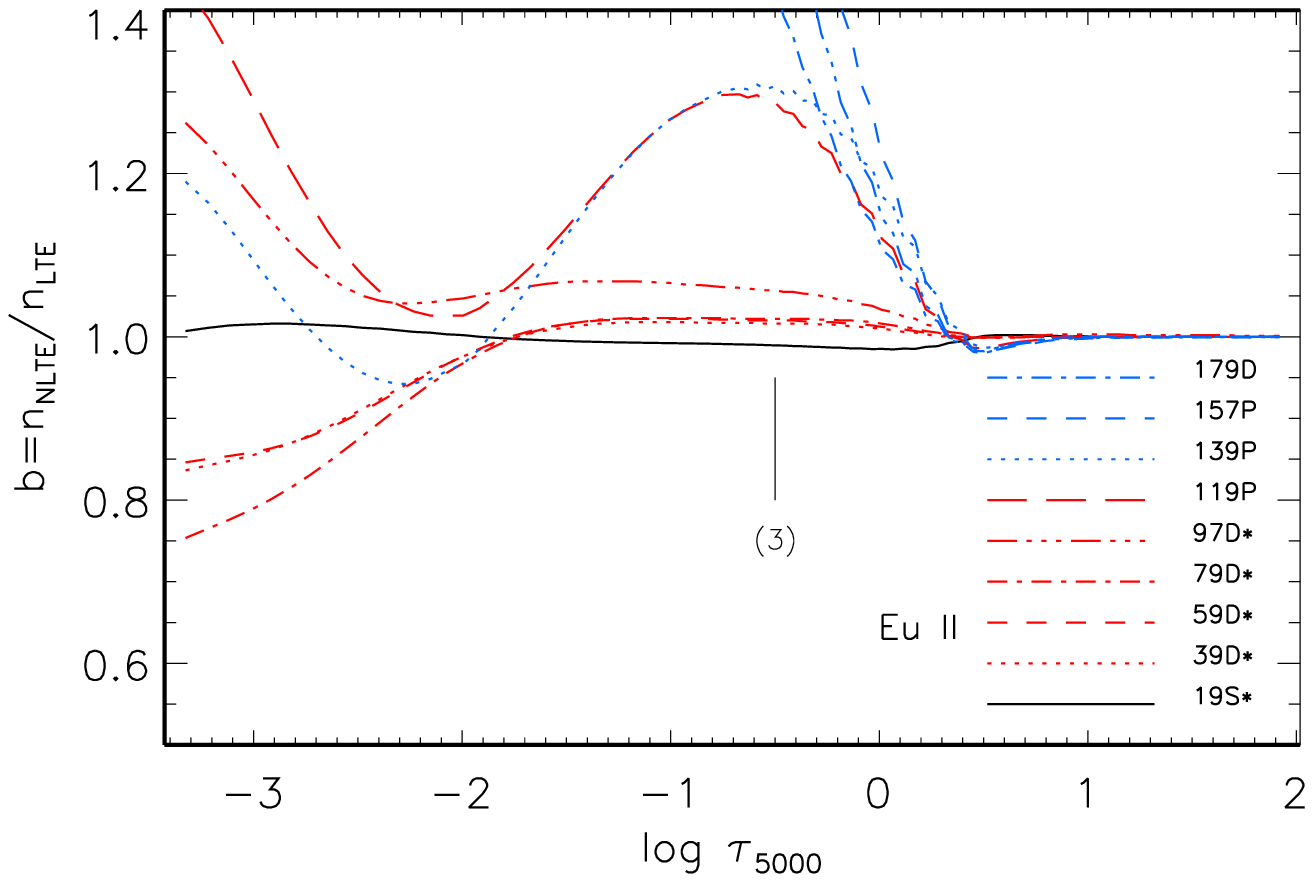}}
\caption[]{Departure coefficients, $b$, for the selected levels of
  \ion{Ba}{ii} (top panel) and \ion{Eu}{ii} (bottom panel) as a function of
  $\log \tau_{5000}$ in the model atmosphere 4800/1.5/$-3.1$. The vertical lines indicate the locations of line core
  formation depths for \ion{Ba}{ii} 4554 (1) and 5853\,\AA\ (2) and \ion{Eu}{ii} 4129\,\AA\ (3). The SE calculations were performed with [Ba/Fe] = 0.7 and [Eu/Fe] = 1.6.} \label{Fig:bf}
\end{figure}

Figure\,\ref{Fig:bf} displays the departure coefficients, $b_i = n_i^{\rm NLTE}/n_i^{\rm LTE}$, for the \ion{Ba}{ii} and \ion{Eu}{ii} levels in the model 4800/1.5/$-3.1$, representing the atmosphere of {\sneden} with effective temperature {\tefft} = 4800~K, surface gravity log~g = 1.5, and iron abundance [Fe/H] = $-3.1$. Here, $n_i^{\rm NLTE}$ and $n_i^{\rm LTE}$ are the SE and TE number densities, respectively. 
 For both \ion{Ba}{ii} and \ion{Eu}{ii}, their ground states (denoted as 6s for \ion{Ba}{ii} and 19S* for \ion{Eu}{ii}) and low-excitation metastable levels (5d32 and 5d52 for \ion{Ba}{ii}; 39D*, 59D*, 79D*, and 97D* for \ion{Eu}{ii}) nearly keep TE level populations, with b $\simeq$ 1, throughout the atmosphere. This is because each of these ions contains the majority of its element. For \ion{Ba}{ii}, the population of the upper level 6p (fine-splitting levels 6p12 and 6p32) of the resonance transition drops below its TE population in the atmospheric layers upwards of $\log \tau_{5000} = -1$, due to photon losses in the resonance lines themselves. The departures from LTE lead to strengthened resonance lines of \ion{Ba}{ii} and negative NLTE abundance corrections $\Delta_{\rm NLTE} = \eps{NLTE}-\eps{LTE}$. For example, $\Delta_{\rm NLTE}$ = $-0.14$~dex for \ion{Ba}{ii} 4554\,\AA. In the layers, where the weakest subordinate line \ion{Ba}{ii} 5853\,\AA\ forms, b(6p) changes its magnitude from above 1 to below 1, resulting in small NLTE effects for the line, with $\Delta_{\rm NLTE}$ = $-0.02$~dex. 
For \ion{Eu}{ii}, radiative pumping of the transitions from the ground state leads to overpopulation of the excited levels in the line-formation layers, resulting in weakened lines and positive NLTE abundance corrections. For example, $\Delta_{\rm NLTE}$ = 0.08~dex for \ion{Eu}{ii} 4129\,\AA.

 \begin{table*} 
 \caption{\label{Tab:stars} Literature data on stellar parameters of the r-II stars and LTE abundances of Ba and Eu.}
 \centering
 \begin{tabular}{lclccrclcccc}\hline\hline \noalign{\smallskip} 
 \multicolumn{1}{c}{Star} &  \multicolumn{1}{c}{\tefft} & log g & [Fe/H] & $\xi_t$ & \multicolumn{3}{c}{Ba abundance} & & \multicolumn{2}{c}{Eu abundance} & Refs \\
\cline{6-8}
\cline{10-11} 
\noalign{\smallskip}  
  & (K) & & & \kms & $\eps{}$ & $f_{\rm odd}$ & Comments & & $\eps{}$ & N &  \\        
\noalign{\smallskip} \hline \noalign{\smallskip} 
CS\,22183-031  & 5270 & 2.8 & $-2.93$ & 1.2 & $-0.33$ & 0.72 & \ion{Ba}{ii} 4554,4934\AA & & $-1.22$ & 3 & 1  \\
CS\,22892-052  & 4800 & 1.5 & $-3.1$ & 1.95 & 0.02 & 0.52 & \ion{Ba}{ii}, 8 lines & & $-0.95$ &  8 &  2  \\
CS\,22953-003 & 5100 & 2.3 & $-2.84$ & 1.7 & $-0.18$ & 0.52 & \ion{Ba}{ii}, 5 lines & & $-1.27$ & 1 &  3 \\
CS\,29497-004 & 5090 & 2.4 & $-2.64$ & 1.6 &  0.50 &  0.72 & \ion{Ba}{ii} 4554,4934\AA & & $-0.45$ &  2 &  4 \\
CS\,31078-018 & 5260 & 2.75 & $-2.85$ & 1.5 & $-0.06$ & 0.72 &  \ion{Ba}{ii} 4554,4934,5853\AA & & $-1.17$ & 4 & 5  \\
CS\,31082-001 & 4825 & 1.5 & $-2.9$ & 1.8 & 0.40 & 0.52 & \ion{Ba}{ii}, 6 lines & & $-0.76$ &  9 &  6 \\
HE\,0432-0923  & 5130 & 2.64 & $-3.19$ & 1.5 & $-0.34$ & 0.72 & \ion{Ba}{ii} 4554\AA & & $-1.43$ &  2 &  7 \\
HE\,1219-0312 & 5060 & 2.3 & $-2.96$ & 1.6 & $-0.14$ & 0.72 & \ion{Ba}{ii} 4554,4934\AA & & $-1.06$ &  3 &  8 \\ 
HE\,1523-091  & 4630 & 1.0 & $-2.95$ & 2.6 & 0.28   & - & - & & $-0.62$ & - & 9 \\
HE\,2224+0143 & 5200 & 2.66 & $-2.58$ & 1.7 & 0.13 & 0.72 & \ion{Ba}{ii} 4554\AA & & $-1.02$ &  4 &  7 \\
HE\,2327-5642 & 5050 & 2.34 & $-2.78$ & 1.8 & $-0.30$ & 0.46 & \ion{Ba}{ii} 5853,6141,6496\AA & & $-1.29$ &  4 &  10 \\ 
SDSS\,J2357-0052 & 5000 & 4.8 & $-3.4$ & 0 & $-0.10$ & - & \ion{Ba}{ii} 5853,6141,6496\AA & & $-0.92$ &  3 & 11  \\
\noalign{\smallskip}\hline \noalign{\smallskip}
\multicolumn{12}{l}{Refs: 1 = \citet{Honda2004}; 2 = \citet{Sneden2003}; 3 = \citet{Francois2007}; 4 = \citet{HERESI}; } \\
\multicolumn{12}{l}{5 = \citet{Lai2008}; 6 = \citet{hill2002}; 7 = \citet{HERESII}; 8 = \citet{HE1219};} \\
\multicolumn{12}{l}{9 = \citet{Sneden2008}; 10 = \citet{HE2327}; 11 = \citet{Aoki2010}.} \\ 
\end{tabular}
\end{table*}

The departures from LTE depend on stellar parameters and the line under investigation. Most r-II stars are VMP cool giants, with the exception of a VMP dwarf discovered by \citet{Aoki2010}. Barium and europium are represented in their spectra mostly in the resonance and low-excitation lines of the majority species, \ion{Ba}{ii} and \ion{Eu}{ii}. We performed NLTE calculations for \ion{Ba}{ii} and \ion{Eu}{ii} in a grid of model atmospheres with {\tefft} = 4500-5250~K, log~g = 1.0-2.75 and log~g = 4.8, [Fe/H] = $-3$ to $-2$, and element abundances characteristic of the r-II stars. In the SE computations, inelastic collisions with neutral hydrogen atoms were taken into account using the classical \citet{Drawin1968,Drawin1969} formalism with a scaling factor of \kH\ = 0.01 for \ion{Ba}{ii} and \kH\ = 0.1 for \ion{Eu}{ii}. This choice is based on our previous analyses of solar and stellar lines of these chemical species \citep{mash_eu}. 
The investigated lines together with their atomic data are listed in Table\,\ref{Tab:lines} (online material).
The results are presented as the LTE equivalent widths, EW (m\AA), and NLTE abundance corrections, $\Delta_{\rm NLTE}$, in Table\,\ref{Tab:ba_nlte} (online material) for lines of \ion{Ba}{ii} and Table\,\ref{Tab:eu_nlte} (online material) for lines of \ion{Eu}{ii}. 

As expected, the NLTE effects for \ion{Ba}{ii} and \ion{Eu}{ii} are minor in the atmosphere of the VMP dwarf (5010/4.8/$-3.4$), with $\Delta_{\rm NLTE} \le$ 0.03~dex in absolute value, and they grow towards lower gravity. For the VMP giant models, $\Delta_{\rm NLTE}$ is negative for the \ion{Ba}{ii} 4554, 4934\,\AA\ resonance lines and low-excitation triplet lines at 5853, 6141, and 6496\,\AA.
 Positive NLTE corrections are predicted for the \ion{Ba}{ii} 3891 and 4130\,\AA\ lines arising from the 6p sublevels. As shown in our previous study \citep{Mashonkina1999}, the departures from LTE for \ion{Ba}{ii} depend on not only stellar parameters, but also the element abundance itself. As can be seen in Table\,\ref{Tab:ba_nlte} (online material), this effect is, in particular, notable for \ion{Ba}{ii} 5853, 6141, 6496\,\AA. For example, in the model atmosphere 4800/1.5/$-3$, the NLTE corrections of these lines reduce by 0.11-0.13~dex, in absolute value, when moving to a 0.4~dex lower Ba abundance. For lines of \ion{Eu}{ii}, $\Delta_{\rm NLTE}$ is positive and ranges between 0.03 and 0.2~dex. 

\subsection{HFS effects}\label{Sect:hfs}

Barium and europium are represented in nature by several isotopes (see Table\,\ref{Tab:ba_isotope} for isotopic abundances of the meteoritic matter).
In the odd-atomic mass isotopes, nucleon-electron spin interactions lead to hyper-fine splitting of the energy levels, resulting in absorption lines divided into multiple components. 
 The existence of isotopic splitting (IS) and/or HFS structure makes the line broader allowing more energy to be absorbed.
Without accounting properly for IS and/or HFS structure,
abundances determined from the lines sensitive to these effects can be
severely overestimated. 

Prominent \ion{Eu}{ii} lines have very broad HFS and IS structure. For example, \ion{Eu}{ii} 4129\,\AA\ consists of, in total, 32 components with 180\,m\AA\ wide patterns. Accurate HFS data for lines of \ion{Eu}{ii} were provided by \citet{Lawler_eu}. 
For the isotope abundance ratio \iso{151}{Eu} : \iso{153}{Eu}, most studies of the r-II stars used either the meteoritic ratio 47.8 : 52.2 (Table\,\ref{Tab:ba_isotope}) or 50 : 50. Both choices are justified, because a pure $r$-process Eu isotope mixture is expected to be very similar to the Solar System one, as shown in Table\,\ref{Tab:rprocess_isotope}. Furthermore, there is observational evidence for equal abundance fractions of 
\iso{151}{Eu} and \iso{153}{Eu} in the VMP stars, including the r-II star {\sneden}, as reported by \citet{Sneden_eu_isotope}. Therefore, Eu abundances of the r-II stars available in the literature do not need a revision for the HFS effect.

 \begin{table} 
 \caption{\label{Tab:rprocess_isotope} Ba and Eu isotope abundance fractions (\%) in the $r$-process.}
 \centering
 \begin{tabular}{llllccll}
\hline\hline \noalign{\smallskip} 
  & \iso{135}{Ba} & \iso{137}{Ba} & \iso{138}{Ba} & & & \iso{151}{Eu} & \iso{153}{Eu} \\
\noalign{\smallskip} \hline \noalign{\smallskip} 
Sneden96 & 40.2 & 12    & 47.8 & & & 47 & 53  \\
McW98    &  40  & 32    & 28   & & & - & - \\
T99 &  24  & 22   & 54   & & & 47.4 & 52.6 \\
B14 & 32  & 28.1  & 39.9 & & & 47.8 & 52.2 \\
A99      &  26  & 20    & 54   & & & 47.4 & 52.6 \\
B11      &  36.8 & 29.3 & 33.9 & & & 47.8 & 52.2 \\
WP       &  29.2 & 14.6 & 56.2 & & & -    & - \\
\noalign{\smallskip}\hline \noalign{\smallskip}
\multicolumn{8}{l}{Sneden96 = \citet{sneden1996}, McW98 = \citet{McWilliam1998},} \\
\multicolumn{8}{l}{T99 = \citet{Travaglio1999}, B14 = \citet{2014arXiv1403.1764B},} \\
\multicolumn{8}{l}{A99 = \citet{Arlandini1999}, B11 = \citet{Bisterzo2011}, } \\
\multicolumn{8}{l}{WP = \citet[][log~n$_n$ = 25]{Kratz2007}.} \\
\end{tabular}
\end{table}

Barium is a different case. Its isotope mixture is very different in the SS matter and the $r$-process, and different studies predicted different isotope abundance fractions for pure $r$-process nucleosynthesis (Table\,\ref{Tab:rprocess_isotope}). The subordinate lines of \ion{Ba}{ii} including prominent 5853, 6141, and 6496\,{\AA} lines are almost free of HFS effects. According to our estimate for
\ion{Ba}{ii} 6496\,{\AA} in the 4500/1.5/$-3$ model with [Ba/Fe] = 1, neglecting HFS makes a difference in abundance of no
more than 0.01\,dex. 

However, about half of the studies of the r-II stars are based on using the 
resonance lines of \ion{Ba}{ii}, which are strongly affected by HFS. For example, \ion{Ba}{ii} 4554\,{\AA} consists of 15 components spread over 58\,m\AA. Three of them are produced by the even-A (relative atomic mass) isotopes, and the isotopic shifts are small, at the level of 1\,m\AA\ \citep{1964CaJPh..42..918K}. Each of the two odd-A isotopes, \iso{135}{Ba} and \iso{137}{Ba}, produce six HFS components disposed by two groups, very similarly for both isotopes. For \ion{Ba}{ii} 4554 and 4934\,{\AA}, most abundance analyses are based on using the HFS patterns published by \citet{McWilliam1998}, which were calculated using the HFS constants from \citet{Brix1952}. Having reviewed the literature for more recent data, \citet{Mashonkina1999} found that the use of new HFS constants of \iso{135}{Ba} and \iso{137}{Ba} from
 \citet{Blatt_hfs137} and \citet{Becker_hfs135} leads to a negligible change in the wavelength
separations by no more than 1\%\, for the HFS components of \ion{Ba}{ii} 4554\,{\AA}. In order to have common input data with other
authors that makes a direct comparison of the results possible, in
this study all calculations of the \ion{Ba}{ii} resonance lines were performed using the wavelength separations and relative intensities of the components from \citet{McWilliam1998}. Table\,\ref{Tab:cs22892} illustrates the effect of using different Ba isotope mixtures on LTE and NLTE abundances derived from the \ion{Ba}{ii} resonance lines in the r-II stars {\sneden} and {\cs}. The difference in NLTE abundance between using different predictions for the $r$-process can be up to 0.14~dex.

\begin{table}
 \caption{\label{Tab:cs22892} Barium LTE and NLTE abundances, $\eps{}$, of {\sneden} and {\cs} for different Ba isotope mixtures.}
 \centering
 \begin{tabular}{lrrrcrr}\hline\hline \noalign{\smallskip} 
\multicolumn{1}{c}{$\lambda$} & \multicolumn{3}{c}{{\sneden}$^1$} & & \multicolumn{2}{c}{{\cs}$^1$} \\
\cline{2-4}
\cline{6-7}
\multicolumn{1}{c}{(\AA)} & EW$^2$ & LTE  & NLTE$^3$ & & LTE$^4$  & NLTE$^{4,5}$  \\
\noalign{\smallskip} \hline \noalign{\smallskip} 
\multicolumn{2}{c}{Ba isotopes:} & \multicolumn{5}{c}{solar mixture, L2009$^6$} \\
\ion{Ba}{ii} 4554$^7$ &  177 & 0.11 & 0.00 & & 0.56 & 0.49 \\  
\ion{Ba}{ii} 4934 &  180 & 0.33 & 0.06 & & 0.94 & 0.63 \\
                  & \multicolumn{6}{c}{$r$-process, \citet{sneden1996}} \\
\ion{Ba}{ii} 4554 &  177 & $-$0.05 & $-$0.18 & & 0.47 & 0.36 \\  
\ion{Ba}{ii} 4934 &  180 & 0.08    & $-$0.22 & & 0.75 & 0.35 \\
                  & \multicolumn{6}{c}{$r$-process, \citet{McWilliam1998}} \\
\ion{Ba}{ii} 4554 &  177 & $-$0.13 & $-$0.28 & & 0.42 & 0.29 \\  
\ion{Ba}{ii} 4934 &  180 & $-$0.02 & $-$0.33 & & 0.66 & 0.23 \\
                  & \multicolumn{6}{c}{$r$-process, \citet{Arlandini1999}} \\
\ion{Ba}{ii} 4554 &  177 & $-$0.03 & $-$0.17 & & 0.48 & 0.37 \\  
\ion{Ba}{ii} 4934 &  180 & 0.07    & $-$0.19 & & 0.56 & 0.29 \\
\multicolumn{2}{l}{mean} & 0.02 & $-$0.18 & & 0.52 & 0.33 \\
                  & \multicolumn{6}{c}{$r$-process, \citet{Bisterzo2011}} \\
\ion{Ba}{ii} 4554 &  177 & $-$0.11 & $-$0.26 & & 0.43 & 0.31 \\  
\ion{Ba}{ii} 4934 &  180 & 0.01    & $-$0.30 & & 0.68 & 0.26 \\
\multicolumn{2}{l}{mean} & $-$0.05 & $-$0.28 & & 0.56 & 0.28 \\
                  &        &        & & & &       \\
\ion{Ba}{ii} 5853 &   72 & $-$0.10 & $-$0.13 & & &   \\
\ion{Ba}{ii} 6141 &  119 & 0.03    & $-$0.17 & & &   \\
\ion{Ba}{ii} 6496 &  117 & 0.11    & $-$0.16 & & &   \\
\multicolumn{2}{l}{mean} & 0.02 & $-$0.15  & & &   \\
\multicolumn{2}{l}{ } & $\pm$0.11 & $\pm$0.02  & & &   \\
\noalign{\smallskip} \hline \noalign{\smallskip} 
\multicolumn{7}{l}{Notes. $^1$ Stellar parameters from Table\,\ref{Tab:stars}, } \\
\multicolumn{7}{l}{ $^2$ equivalent widths (m\AA) from \citet{sneden1996}, } \\
\multicolumn{7}{l}{ $^3$ $\eps{Ba}$ = $-0.11$ in the SE calculations, } \\
\multicolumn{7}{l}{ $^4$ from line profile fitting, $^5$ $\eps{Ba}$ = 0.24 in the SE calculations,} \\
\multicolumn{7}{l}{ $^6$ \citet{Lodders2009},} \\
\multicolumn{7}{l}{ $^7$ line data from Table\,\ref{Tab:lines} (online material).}  \\
\end{tabular}
\end{table}

From observations, the total fractional abundance of \iso{135}{Ba} and \iso{137}{Ba}, $f_\mathrm{odd}$, can only be derived because these isotopes have very similar HFS. 
In the SS matter, $f_\mathrm{odd} = 0.18$. For the $r$-process, different studies predicted $f_\mathrm{odd}$ = 0.46 to 0.72 (Table\,\ref{Tab:rprocess_isotope}). Stellar Ba isotopic fractions were determined in few studies. \citet{Magain1993} have suggested a method based on measuring the broadening of the \ion{Ba}{ii} 4554\,\AA\ line in a very high-quality observed spectrum (resolving power $R = \lambda/\Delta\lambda \simeq 100\,000$ and signal-to-noise ratio $S/N \simeq 180$) of the nearby halo star HD\,140283. Even higher $R$ and $S/N$ spectra of this star were then used by \citet[][$R\simeq 100\,000$, $S/N\simeq 400$]{Magain1995}, \citet[][$R\simeq 200\,000$, $S/N\simeq 550$]{2002MNRAS.335..325L}, and \citet[][$S/N\simeq 1\,110$]{Gallagher2010}, but the obtained results were contrasting. \citet{Magain1995} and \citet{Gallagher2010} found that the odd-A isotopes constitute a minor fraction of the total Ba in HD\,140283, with $f_\mathrm{odd}$ = 0.08 and 0.02, respectively, suggesting a pure $s-$process production of barium. It is worth noting that \citet[][stellar model]{Arlandini1999} predicted $f_\mathrm{odd}^s = 0.11$ for the $s-$process. \citet{2002MNRAS.335..325L} obtained $f_\mathrm{odd} = 0.30\pm0.21$ which covers the full range of possibilities from a pure $s-$ to a pure $r-$process production of Ba. Completely disappointing results were presented by \citet{Gallagher2012} for three metal-poor stars, with $f_\mathrm{odd}$ = $-0.12\pm0.07$, $-0.02\pm0.09$, and $-0.05\pm0.11$ for HD\,122563, HD\,88609, and HD\,84937, respectively. The fractions $f_\mathrm{odd}$ = $0.08\pm0.08$ and $0.18\pm0.08$ measured in another two halo stars BD~+26$^\circ$3578 and BD$-04^\circ$3208 have, at least, a physical sense. \citet{Gallagher2012} concluded that
``it is much more likely that the symmetric
1D LTE techniques used in this investigation are inadequate and
improvements to isotopic ratio analysis need to be made.''
The data cited above for HD\,140283 were also based on a LTE analysis using 1D model atmospheres. The three-dimensional (3D) hydrodynamical model and LTE assumption were only applied by \citet{2009PASA...26..330C}. With the observed spectrum from \citet{2002MNRAS.335..325L}, they obtained $f_\mathrm{odd} = 0.15\pm0.12$ for HD140283.

A different method based on abundance comparisons between the subordinate and resonance lines of \ion{Ba}{ii} was used in our earlier studies. The subordinate lines provide the total abundance of Ba in a star. The element abundance is then deduced from the resonance lines for
various isotopic mixtures, the proportion of the odd isotopes
being changed until agreement with the subordinate lines is obtained. 
Table\,\ref{Tab:cs22892} can serve to illustrate this method. For {\sneden}, consistent abundances from different lines are achieved with $f_\mathrm{odd} \simeq 0.45$. However, one needs to be very cautious when applying such an 
approach to a star in which the resonance lines of \ion{Ba}{ii} are saturated and very sensitive to variation in microturbulence velocity $\xi_{\rm t}$. For {\sneden}, $\Delta\xi_{\rm t} = 0.2$~\kms\ produces a change in abundance of 0.15~dex that is comparable with the HFS effect. Taking advantage of NLTE line formation calculations of \ion{Ba}{ii}, \citet{Mashonkina2006} and \citet{Mashonkina2008} derived $f_\mathrm{odd}$-values for three nearby halo stars and 15 thick disk stars with well determined stellar parameters. The two halo stars, HD\,103095 ($f_\mathrm{odd}$ = 0.42$\pm$0.06) and HD\,84937 ($f_\mathrm{odd}$ = 0.43$\pm$0.14), and the thick disk stars (the mean $f_\mathrm{odd}$ = 0.33$\pm$0.04) reveal higher fractions of the odd isotopes of Ba compared with that for the solar Ba isotope mixture. For HD\,122563, the obtained $f_\mathrm{odd}$ = 0.22$\pm$0.15 is close to the SS one. 

The observational evidence summarised above indicates that
stellar $f_\mathrm{odd}$ is related to the $r$-process abundances of the star. Indeed, all the stars with high $f_\mathrm{odd}$-values, i.e., HD\,103095, HD\,84937, and the investigated thick disk stars, reveal an enhancement of europium relative to iron, with [Eu/Fe] = 0.24 to 0.70 \citep{Mashonkina2001sr,Mashonkina2008}. Analysis of the r-II star {\LyudmilasStar}, with [Eu/Fe] = 0.98, also favored a high fraction of the odd isotopes of Ba, with $f_\mathrm{odd} \simeq 0.5$ \citep{HE2327}. In contrast, both stars with low $f_\mathrm{odd}$ are Eu-poor. For HD\,122563, [Eu/Fe] = $-0.51$ / $-0.63$ \citep[NLTE / LTE,][]{Mashonkina2008}, and only an upper limit, [Eu/Fe] $< -0.2$, was obtained for HD\,140283 \citep{Gallagher2010}.

\section{Barium and europium abundances of the r-II stars}\label{Sect:stars}

In this study, the NLTE abundances of Ba and Eu were obtained for 8 of the 12 r-II stars currently known. HE\,1523-091 was not included in our analysis, because there is no list of the used lines in \citet{Sneden2008}. CS\,31078-018 was not included, too, because we could not reproduce the LTE element abundances published by \citet{Lai2008}, when employing their online material for observed equivalent widths of the \ion{Ba}{ii} and \ion{Eu}{ii} lines. We also did not use HE\,0432-0923 and HE\,2224+0143, with stellar parameters and element abundances based on an automated line profile analysis of moderate-resolution ($R \simeq$ 20\,000) ``snapshot'' spectra (Paper~II).

For the investigated stars, the NLTE calculations were performed with 
stellar parameters taken from the literature. They are listed in Table\,\ref{Tab:stars}. The element abundances were derived using the line lists from the original papers. An exception is HE\,1219-0312, where we employed an extended list of the \ion{Ba}{ii} and \ion{Eu}{ii} lines from \citet{HE2327}. For the final abundances of Ba, we preferred to use the subordinate lines of \ion{Ba}{ii}, which are nearly free of the HFS effects. In cases where only the resonance lines were available, homogeneous abundances were calculated using a common mixture of the Ba isotopes. Five different isotope mixtures were checked for {\sneden} and {\cs} (Table\,\ref{Tab:cs22892}) and two, as predicted by A99 (and also T99) and B11 for the $r$-process, for the remaining stars.  
For Eu in all the stars except {\cs}, HE\,1219-0312, and HE\,2327-5642, the NLTE calculations were performed with published LTE abundances, and the final NLTE abundances were obtained by applying the NLTE corrections. 
Table\,\ref{Tab:ba_eu} presents the obtained results for the option of using the A99 (T99) $r$-process. Below, we comment on the individual stars.

 \begin{table*} 
 \caption{\label{Tab:ba_eu} Obtained LTE and NLTE abundances, $\eps{}$, and abundance ratios log(Ba/Eu) of the $r$-process enhanced stars.}
 \centering
 \begin{tabular}{lllccllcl}\hline\hline \noalign{\smallskip} 
 &  \multicolumn{3}{c}{LTE} & & \multicolumn{3}{c}{NLTE} & \multicolumn{1}{c}{Notes} \\
\cline{2-4}
\cline{6-8}
\noalign{\smallskip}  
 \multicolumn{1}{c}{Star} & \multicolumn{1}{c}{Ba} & \multicolumn{1}{c}{Eu} & Ba/Eu & & \multicolumn{1}{c}{Ba} & \multicolumn{1}{c}{Eu} & Ba/Eu & \\        
\noalign{\smallskip} \hline \noalign{\smallskip} 
 & \multicolumn{8}{l}{Ba abundance from the \ion{Ba}{ii} subordinate lines} \\
{\sneden}  &  ~~0.02$\pm0.11$  & $-0.95$$\pm$0.03 &  0.97  & & $-0.15$$\pm$0.02 &  $-0.85$$\pm$0.03 & 0.70 & (a) \\
HE\,1219-0312  & $-0.06$$\pm$0.03 & $-0.99$$\pm$0.03 &  0.93  & & $-0.15$$\pm$0.10 &  $-0.88$$\pm$0.01 & 0.73 & (b) \\ 
HE\,2327-5642  &  $-0.30$$\pm$0.03 & $-1.29$$\pm$0.02 &  0.99 & & $-0.30$$\pm$0.06 &  $-1.12$$\pm$0.04 & 0.82 & (c) \\ 
SDSS\,J2357-0052 & $-0.10$$\pm$0.09 & $-0.92$$\pm$0.19 &  0.82 & & $-0.09$$\pm$0.09 &  $-0.89$$\pm$0.19 & 0.80 & (d) \\
 & \multicolumn{8}{l}{Ba abundance from the subordinate and resonance lines, $f_{odd} = 0.5$} \\
CS\,22953-003 & $-0.22$ & $-1.27$ &  1.05   & & $-0.23$ &  $-1.12$ & 0.89 & (e) \\
{\cayrel} &  ~~0.39	 & $-0.76$$\pm$0.11 &  1.07   & & ~~0.05    &  $-0.70$$\pm$0.11 & 0.75 & (e) \\
 & \multicolumn{8}{l}{Ba abundance from the resonance lines, $f_{odd} = 0.5$} \\
CS\,22183-031 & $-0.21$$\pm$0.15 & $-1.22$$\pm$0.08 &  1.01   & & $-0.35$$\pm$0.15 &  $-1.08$$\pm$0.08 & 0.73 & (d) \\
{\cs}         &  ~~0.50$\pm$0.06 & $-0.54$$\pm$0.07 &  1.04   & & ~~0.30$\pm$0.06  &  $-0.46$$\pm$0.05 & 0.76 & (b) \\
\cline{1-1}	 
\cline{4-4}	 
\cline{8-8}	 
mean           &         &         &  0.98   & &         &          & 0.77 & \\
\noalign{\smallskip}\hline \noalign{\smallskip}
\multicolumn{9}{l}{Notes. (a) Ba: EW, \citet{sneden1996}; Eu: $\Delta_{\rm NLTE}$, (b) syn, VLT/UVES spectrum, } \\
\multicolumn{9}{l}{ (c) \citet{HE2327}, (d) $\Delta_{\rm NLTE}$, (e) Ba: \citet{Andri2009}; Eu: $\Delta_{\rm NLTE}$. } \\
\end{tabular}
\end{table*}

\underline{{\cs} and HE\,1219-0312}. The LTE and NLTE abundances of Ba and Eu were derived from line profile fitting of the UVES/VLT spectra taken from the original papers of \citet{HERESI} and \citet{HE1219}, respectively. For Ba in HE\,1219-0312, the final abundance is based on the subordinate lines. The spectrum of {\cs} did not cover the 5853-6496\,\AA\ range, and the Ba abundances were determined from the resonance lines for various isotope mixtures (Table\,\ref{Tab:cs22892}).

\underline{{\sneden}}. For Ba, we used the equivalent widths published by \citet{sneden1996} and the final abundance is based on the three subordinate lines (Table\,\ref{Tab:cs22892}). We wish to note a dramatic reduction in the statistical abundance error from 0.11 to 0.02~dex, when moving from LTE to NLTE. This also holds for the two resonance lines. These results favor the NLTE line formation for \ion{Ba}{ii}.
    
\underline{CS\,22953-003 and CS\,31082-001}. The NLTE abundances of Ba were taken from \citet{Andri2009}, where they were derived from \ion{Ba}{ii} 5853, 6496, and 4554\,\AA\ using $f_\mathrm{odd} = 0.5$.

\underline{HE\,2327-5642}. The NLTE abundances of Ba and Eu were determined in our earlier study \citep{HE2327}, using only the \ion{Ba}{ii} subordinate lines for Ba.

\underline{CS\,22183-031 and SDSS\,J2357-0052}. The NLTE abundances of Ba and Eu were calculated by applying NLTE corrections computed for individual spectral lines. SDSS\,J2357-0052 is the only dwarf among the r-II stars, and it reveals small departures from LTE. 
For CS\,22183-031, the published LTE abundance of Ba (Table\,\ref{Tab:stars}) was transformed to $f_\mathrm{odd} = 0.5$ by adding a HFS correction of +0.12~dex.

\begin{figure}  
  \resizebox{88mm}{!}{\includegraphics{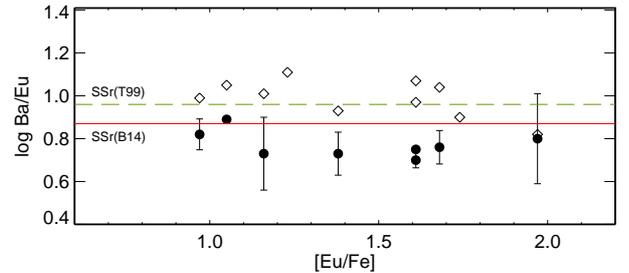}}
  \caption{\label{Fig:r2} The LTE (rombs) and NLTE (filled circles) Ba/Eu abundance ratios of the r-II stars. The Ba abundance from the \ion{Ba}{ii} resonance lines was derived using $f_\mathrm{odd} = 0.5$.  
The continuous and dashed lines indicate the SSr ratios log(Ba/Eu)$_r$ = 0.87 and 0.96, as predicted by B14 and T99}, respectively.
\end{figure}

The obtained Ba/Eu abundance ratios are displayed in Fig.\,\ref{Fig:r2}. 
In LTE, the mean from 10 stars is log(Ba/Eu) = 0.99$\pm$0.09. Here, we include CS\,31078-018 and HE\,1523-091, with the original LTE ratios log(Ba/Eu) = 1.11 \citep{Lai2008} and 0.90 \citep{Sneden2008}. In NLTE, log(Ba/Eu) = 0.78$\pm$0.06 and 0.75$\pm$0.07 from 8 stars, when using the $r$-process Ba isotope mixtures from A99 (T99) and B11, respectively, for the four stars with Ba abundances determined from the \ion{Ba}{ii} resonance lines.
The NLTE Ba/Eu abundance ratios of the r-II stars support the recent predictions of \citet{Bisterzo2011} and \citet{2014arXiv1403.1764B} for the solar $r$-residuals and also the HEW $r$-process model of \citet{Farouqi2010}.

In this study, we did not consider the effects on the Ba and Eu abundances of the use of 3D hydrodynamical model atmospheres. Both elements are observed in the r-II stars in the lines of their majority species, and the detected lines arise from either the ground or low-excitation levels. \citet{2013A&A...559A.102D} predicted that the (3D-1D) abundance corrections are small and very similar in magnitude for the $E_{\rm exc}$ = 0 lines of \ion{Ba}{ii} and \ion{Eu}{ii} in the MP cool giant models. For example, (3D-1D) $= -0.05$~dex and $-0.04$~dex for the resonance lines of \ion{Ba}{ii} and \ion{Eu}{ii}, respectively, in the 5020/2.5/$-3$ model.

\section{Conclusions}\label{Sect:Conclusions}

In this study, the Ba and Eu abundances of the eight r-II stars were revised by taking departures from LTE for lines of \ion{Ba}{ii} and \ion{Eu}{ii} into account, and accounting for HFS affecting the \ion{Ba}{ii} resonance lines with a common Ba isotope mixture. For most of the r-II stars, NLTE
 leads to a lower Ba, but a higher Eu abundance. Therefore, the Ba/Eu
 abundance ratios decrease on average by 0.21~dex when 
moving from LTE to NLTE
We conclude that an adequate line-formation modelling for heavy elements is important for abundance comparisons between VMP stars, and in particular, giants.

For stellar Ba abundance determinations, we recommend to use the subordinate lines of \ion{Ba}{ii}, which are nearly free of the HFS effects. 
At present, there is no consensus on the Ba isotope abundance fractions in the $r$-process, and abundances derived from the \ion{Ba}{ii} resonance lines vary by 0.15~dex, when applying different $r$-process models.
We suspect that the total fractional abundance of the odd-A isotopes of Ba in old Galactic stars 
 is related to the $r$-process abundances of the star. The observational evidence indicates that all the stars with high $f_\mathrm{odd}$-values, i.e., HD\,103095, HD\,84937, {\LyudmilasStar}, and the selected thick disk stars, are $r$-process enhanced, with [Eu/Fe] = 0.24 to 0.70 \citep{Mashonkina2006,Mashonkina2008,HE2327}. In contrast, both stars with low $f_\mathrm{odd}$, i.e., HD\,122563 and HD\,140283, are Eu-poor, with [Eu/Fe] = $-0.51$ 
 \citep[][NLTE]{Mashonkina2008} and [Eu/Fe] $< -0.2$ \citep{Gallagher2010}, respectively.

With the improved Ba and Eu abundances of the r-II stars, we constrain a pure $r$-process Ba/Eu abundance ratio to be
log(Ba/Eu)$_r$ = 0.78$\pm$0.06. The obtained results support the solar $r$-residuals based on the chemical evolution calculations of \citet{Bisterzo2011} and \citet{2014arXiv1403.1764B}, and also the HEW $r$-process model by \citet{Farouqi2010}.

For further constraining the $r$-process models, it would be important to determine Ba isotopic fractions of the r-II stars. 
This is a challenge for theory as well as observations.

\begin{acknowledgements}
  
This work was supported by Sonderforschungsbereich SFB
 881 'The Milky Way System' (subprojects A4 and A5) of 
the German   Research Foundation (DFG).
L.M. was supported by the RF President with a grant on Leading Scientific Schools 3620.2014.2 and the Swiss National Science Foundation (SCOPES project No.~IZ73Z0-128180/1).
We made use the NIST and VALD databases.

\end{acknowledgements}
\bibliography{ml_pb,nlte,mp_stars,references}
\bibliographystyle{aa}

\Online

\begin{table}
 \caption{\label{Tab:lines} Atomic data for the investigated lines. $\Gamma_6$ corresponds to 10\,000~K. References to the adopted $gf-$values and $\Gamma_6-$values are given in column 4 and 6, respectively.}
 \centering
 \begin{tabular}{rcrccc}\hline\hline \noalign{\smallskip} 
\multicolumn{1}{c}{$\lambda$(\AA)} & $E_{\rm exc}$(eV) & log $gf$ & Refs & log$\Gamma_6$ & Refs \\
\noalign{\smallskip} \hline \noalign{\smallskip} 
\ion{Ba}{ii} 3891.78 & 2.51 & 0.28 & 1 & $-$7.870 & 3 \\
\ion{Ba}{ii} 4130.64 & 2.72 & 0.56 & 1 & $-$7.870 & 3 \\
\ion{Ba}{ii} 4554.03 & 0.00 & 0.17 & 1 & $-$7.732 & 4 \\  
\ion{Ba}{ii} 4934.08 & 0.00 & $-$0.15 & 1 & $-$7.732 & 4 \\
\ion{Ba}{ii} 5853.67 & 0.60 & $-$1.01 & 1 & $-$7.584 & 5 \\
\ion{Ba}{ii} 6141.71 & 0.70 & $-$0.07 & 1 & $-$7.584 & 5 \\
\ion{Ba}{ii} 6496.90 & 0.60 & $-$0.38 & 1 & $-$7.584 & 5 \\
\ion{Eu}{ii} 3724.93 & 0.00 & $-$0.09 & 2 & $-$7.870 & 3 \\	 
\ion{Eu}{ii} 3819.67 & 0.00 &    0.51 & 2 & $-$7.870 & 3 \\
\ion{Eu}{ii} 3907.11 & 0.21 &    0.17 & 2 & $-$7.870 & 3 \\
\ion{Eu}{ii} 3930.50 & 0.21 &    0.27 & 2 & $-$7.870 & 3 \\	
\ion{Eu}{ii} 3971.97 & 0.21 &    0.27 & 2 & $-$7.870 & 3 \\	
\ion{Eu}{ii} 4129.72 & 0.00 &    0.22 & 2 & $-$7.870 & 3 \\	
\ion{Eu}{ii} 4205.02 & 0.00 &    0.21 & 2 & $-$7.870 & 3 \\	
\ion{Eu}{ii} 4435.58 & 0.21 & $-$0.11 & 2 & $-$7.870 & 3 \\	
\ion{Eu}{ii} 4522.58 & 0.21 & $-$0.67 & 2 & $-$7.870 & 3 \\	
\ion{Eu}{ii} 6645.10 & 1.38 &   0.12  & 2 & $-$7.870 & 3 \\       
\noalign{\smallskip}\hline \noalign{\smallskip}
\multicolumn{6}{l}{Refs. 1 = \citet{Gallagher_ba}, 2 = \citet{Lawler_eu}, } \\
\multicolumn{6}{l}{ \ 3 = adopted in this study, 4 = \citet{Mashonkina2006},} \\
\multicolumn{6}{l}{ \ 5 = \citet{1998MNRAS.300..863B}.} \\
\end{tabular}
\end{table}

\longtab{7}{
\begin{longtable}[]{lcrrrrrrrr}   
 \caption{\label{Tab:ba_nlte} Equivalent widths, EW$_{\rm LTE}$(m\AA), and NLTE abundance corrections, $\Delta_{\rm NLTE}$(dex), for lines of \ion{Ba}{ii} in the metal-poor model atmospheres. All the computations were performed with $\xi_{\rm t}$ = 1.8~\kms\ except for the log~g = 4.8 model, where $\xi_{\rm t}$ = 0 was adopted. Everywhere, \kH\ = 0.01. For line data, see Table\,\ref{Tab:lines}. Lines with EW $<$ 5\,m\AA\ are not shown.} \\
\hline\hline \noalign{\smallskip} 
Model & [Ba/Fe] & & \multicolumn{7}{c}{\ion{Ba}{ii} lines, $\lambda$(\AA)} \\
\cline{4-10} \noalign{\smallskip} 
      &         & & 4554 & 4934 & 5853 & 6141 & 6496 & 4130 & 3891 \\
\noalign{\smallskip} \hline \noalign{\smallskip} 
\endfirsthead
\caption{continued.}\\
\hline \noalign{\smallskip} 
Model & [Ba/Fe] & & \multicolumn{7}{c}{\ion{Ba}{ii} lines, $\lambda$(\AA)} \\
\cline{4-10} \noalign{\smallskip} 
      &         & & 4554 & 4934 & 5853 & 6141 & 6496 & 4130 & 3891 \\
\hline  \\   
\endhead
\hline
\endfoot
\hline
\endlastfoot
 4500/1.00/$-$3.00 &  1.0 & 
 EW$_{\rm LTE}$ &   237 &   223 &   106 &   149 &   144 &    28 &    28 \\ 
 & & $\Delta_{\rm NLTE}$ & $-$0.03 & $-$0.09 & $-$0.13 & $-$0.17 & $-$0.24 &  0.41 &  0.34 \\ 
 4500/1.00/$-$3.00 &  0.5 & 
 EW$_{\rm LTE}$ &   187 &   183 &    79 &   119 &   114 &    12 &    12 \\ 
 & & $\Delta_{\rm NLTE}$ & $-$0.04 & $-$0.14 & $-$0.04 & $-$0.13 & $-$0.19 &  0.33 &  0.29 \\ 
 4500/1.00/$-$3.00 &  0.0 & 
 EW$_{\rm LTE}$ &   153 &   152 &    50 &    92 &    86 &     &     \\ 
 & & $\Delta_{\rm NLTE}$ & $-$0.03 & $-$0.11 &  0.04 & $-$0.01 & $-$0.06 &   &   \\ 
 4500/1.50/$-$3.00 &  1.0 & 
 EW$_{\rm LTE}$ &   231 &   215 &    99 &   141 &   135 &    22 &    22 \\ 
 & & $\Delta_{\rm NLTE}$ & $-$0.04 & $-$0.12 & $-$0.13 & $-$0.20 & $-$0.26 &  0.31 &  0.26 \\ 
 4500/1.50/$-$3.00 &  0.5 & 
 EW$_{\rm LTE}$ &   179 &   174 &    71 &   111 &   105 &     8 &     8 \\ 
 & & $\Delta_{\rm NLTE}$ & $-$0.06 & $-$0.16 & $-$0.03 & $-$0.13 & $-$0.18 &  0.26 &  0.23 \\ 
 4500/1.50/$-$3.00 &  0.0 & 
 EW$_{\rm LTE}$ &   145 &   143 &    41 &    83 &    77 &     &     \\ 
 & & $\Delta_{\rm NLTE}$ & $-$0.04 & $-$0.11 &  0.04 &  0.00 & $-$0.04 &   &   \\ 
 4500/2.00/$-$3.00 &  1.0 & 
 EW$_{\rm LTE}$ &   230 &   210 &    92 &   135 &   127 &    16 &    16 \\ 
 & & $\Delta_{\rm NLTE}$ & $-$0.05 & $-$0.13 & $-$0.10 & $-$0.19 & $-$0.24 &  0.24 &  0.20 \\ 
 4500/2.00/$-$3.00 &  0.5 & 
 EW$_{\rm LTE}$ &   174 &   167 &    61 &   103 &    96 &     6 &     6 \\ 
 & & $\Delta_{\rm NLTE}$ & $-$0.07 & $-$0.16 & $-$0.01 & $-$0.11 & $-$0.15 &  0.22 &  0.19 \\ 
 4500/2.00/$-$3.00 &  0.0 & 
 EW$_{\rm LTE}$ &   138 &   134 &    32 &    74 &    66 &     &     \\ 
 & & $\Delta_{\rm NLTE}$ & $-$0.04 & $-$0.09 &  0.04 &  0.01 & $-$0.02 &   &   \\ 
 4750/1.00/$-$3.00 &  1.0 & 
 EW$_{\rm LTE}$ &   201 &   193 &    94 &   132 &   127 &    26 &    25 \\ 
 & & $\Delta_{\rm NLTE}$ & $-$0.11 & $-$0.24 & $-$0.13 & $-$0.27 & $-$0.33 &  0.34 &  0.29 \\ 
 4750/1.00/$-$3.00 &  0.5 & 
 EW$_{\rm LTE}$ &   165 &   163 &    69 &   107 &   101 &    10 &    10 \\ 
 & & $\Delta_{\rm NLTE}$ & $-$0.12 & $-$0.23 & $-$0.01 & $-$0.14 & $-$0.19 &  0.28 &  0.25 \\ 
 4750/1.00/$-$3.00 &  0.0 & 
 EW$_{\rm LTE}$ &   138 &   137 &    39 &    82 &    75 &     &     \\ 
 & & $\Delta_{\rm NLTE}$ & $-$0.08 & $-$0.12 &  0.07 &  0.01 & $-$0.01 &   &   \\ 
 4750/1.50/$-$3.00 &  1.0 & 
 EW$_{\rm LTE}$ &   193 &   185 &    86 &   123 &   117 &    20 &    19 \\ 
 & & $\Delta_{\rm NLTE}$ & $-$0.13 & $-$0.27 & $-$0.12 & $-$0.28 & $-$0.33 &  0.28 &  0.24 \\ 
 4750/1.50/$-$3.00 &  0.5 & 
 EW$_{\rm LTE}$ &   157 &   155 &    59 &    98 &    92 &     8 &     7 \\ 
 & & $\Delta_{\rm NLTE}$ & $-$0.14 & $-$0.23 &  0.00 & $-$0.14 & $-$0.17 &  0.24 &  0.21 \\ 
 4750/1.50/$-$3.00 &  0.0 & 
 EW$_{\rm LTE}$ &   130 &   128 &    31 &    73 &    65 &     &     \\ 
 & & $\Delta_{\rm NLTE}$ & $-$0.08 & $-$0.11 &  0.06 &  0.02 &  0.00 &   &   \\ 
 4750/2.00/$-$3.00 &  1.0 & 
 EW$_{\rm LTE}$ &   189 &   179 &    79 &   117 &   110 &    15 &    14 \\ 
 & & $\Delta_{\rm NLTE}$ & $-$0.13 & $-$0.26 & $-$0.10 & $-$0.26 & $-$0.30 &  0.23 &  0.20 \\ 
 4750/2.00/$-$3.00 &  0.5 & 
 EW$_{\rm LTE}$ &   151 &   147 &    50 &    90 &    84 &     5 &     5 \\ 
 & & $\Delta_{\rm NLTE}$ & $-$0.14 & $-$0.22 &  0.00 & $-$0.12 & $-$0.14 &  0.21 &  0.19 \\ 
 4750/2.00/$-$3.00 &  0.0 & 
 EW$_{\rm LTE}$ &   124 &   119 &    24 &    64 &    55 &     &     \\ 
 & & $\Delta_{\rm NLTE}$ & $-$0.07 & $-$0.09 &  0.05 &  0.03 &  0.01 &   &   \\ 
 5000/2.00/$-$3.00 &  1.0 & 
 EW$_{\rm LTE}$ &   165 &   160 &    68 &   104 &    97 &    13 &    12 \\ 
 & & $\Delta_{\rm NLTE}$ & $-$0.17 & $-$0.29 & $-$0.05 & $-$0.24 & $-$0.27 &  0.23 &  0.20 \\ 
 5000/2.00/$-$3.00 &  0.5 & 
 EW$_{\rm LTE}$ &   136 &   133 &    39 &    80 &    72 &     5 &     \\ 
 & & $\Delta_{\rm NLTE}$ & $-$0.13 & $-$0.17 &  0.05 & $-$0.05 & $-$0.06 &  0.22 &   \\ 
 5000/2.00/$-$3.00 &  0.0 & 
 EW$_{\rm LTE}$ &   112 &   106 &    17 &    53 &    44 &     &     \\ 
 & & $\Delta_{\rm NLTE}$ & $-$0.01 &  0.00 &  0.09 &  0.09 &  0.08 &   &   \\ 
 5000/2.50/$-$3.00 &  1.0 & 
 EW$_{\rm LTE}$ &   160 &   154 &    59 &    97 &    90 &    10 &     9 \\ 
 & & $\Delta_{\rm NLTE}$ & $-$0.16 & $-$0.26 & $-$0.03 & $-$0.20 & $-$0.22 &  0.22 &  0.19 \\ 
 5000/2.50/$-$3.00 &  0.5 & 
 EW$_{\rm LTE}$ &   130 &   125 &    31 &    72 &    63 &     &     \\ 
 & & $\Delta_{\rm NLTE}$ & $-$0.11 & $-$0.13 &  0.04 & $-$0.03 & $-$0.04 &   &   \\ 
 5000/2.50/$-$3.00 &  0.0 & 
 EW$_{\rm LTE}$ &   104 &    96 &    12 &    43 &    35 &     &     \\ 
 & & $\Delta_{\rm NLTE}$ &  0.01 &  0.02 &  0.08 &  0.09 &  0.07 &   &   \\ 
 4500/1.00/$-$2.00 &  0.5 & 
 EW$_{\rm LTE}$ &   284 &   260 &   125 &   174 &   167 &    46 &    46 \\ 
 & & $\Delta_{\rm NLTE}$ & $-$0.01 & $-$0.05 & $-$0.12 & $-$0.12 & $-$0.17 &  0.30 &  0.25 \\ 
 4500/1.00/$-$2.00 &  0.5 & 
 EW$_{\rm LTE}$ &   216 &   208 &    98 &   140 &   134 &    24 &    23 \\ 
 & & $\Delta_{\rm NLTE}$ & $-$0.03 & $-$0.08 & $-$0.09 & $-$0.14 & $-$0.18 &  0.27 &  0.23 \\ 
 4500/1.50/$-$2.00 &  0.5 & 
 EW$_{\rm LTE}$ &   267 &   244 &   115 &   161 &   154 &    38 &    37 \\ 
 & & $\Delta_{\rm NLTE}$ & $-$0.03 & $-$0.07 & $-$0.14 & $-$0.16 & $-$0.21 &  0.25 &  0.20 \\ 
 4500/1.50/$-$2.00 &  0.0 & 
 EW$_{\rm LTE}$ &   203 &   196 &    87 &   128 &   122 &    18 &    17 \\ 
 & & $\Delta_{\rm NLTE}$ & $-$0.04 & $-$0.11 & $-$0.08 & $-$0.16 & $-$0.19 &  0.21 &  0.18 \\ 
 4500/2.00/$-$2.00 &  0.5 & 
 EW$_{\rm LTE}$ &   257 &   233 &   105 &   151 &   142 &    30 &    29 \\ 
 & & $\Delta_{\rm NLTE}$ & $-$0.04 & $-$0.08 & $-$0.12 & $-$0.17 & $-$0.21 &  0.20 &  0.16 \\ 
 4500/2.00/$-$2.00 &  0.0 & 
 EW$_{\rm LTE}$ &   194 &   185 &    77 &   117 &   110 &    12 &    12 \\ 
 & & $\Delta_{\rm NLTE}$ & $-$0.05 & $-$0.12 & $-$0.05 & $-$0.15 & $-$0.17 &  0.17 &  0.14 \\ 
 4750/1.00/$-$2.00 &  0.5 & 
 EW$_{\rm LTE}$ &   254 &   237 &   117 &   162 &   155 &    47 &    45 \\ 
 & & $\Delta_{\rm NLTE}$ & $-$0.04 & $-$0.09 & $-$0.17 & $-$0.20 & $-$0.27 &  0.32 &  0.27 \\ 
 4750/1.00/$-$2.00 &  0.0 & 
 EW$_{\rm LTE}$ &   199 &   194 &    92 &   131 &   126 &    24 &    23 \\ 
 & & $\Delta_{\rm NLTE}$ & $-$0.06 & $-$0.15 & $-$0.09 & $-$0.20 & $-$0.25 &  0.27 &  0.24 \\ 
 4750/1.50/$-$2.00 &  0.5 & 
 EW$_{\rm LTE}$ &   239 &   223 &   108 &   150 &   143 &    39 &    37 \\ 
 & & $\Delta_{\rm NLTE}$ & $-$0.05 & $-$0.13 & $-$0.17 & $-$0.24 & $-$0.29 &  0.27 &  0.22 \\ 
 4750/1.50/$-$2.00 &  0.0 & 
 EW$_{\rm LTE}$ &   188 &   183 &    82 &   120 &   114 &    18 &    17 \\ 
 & & $\Delta_{\rm NLTE}$ & $-$0.08 & $-$0.18 & $-$0.08 & $-$0.20 & $-$0.24 &  0.22 &  0.19 \\ 
 4750/2.00/$-$2.00 &  0.5 & 
 EW$_{\rm LTE}$ &   229 &   212 &    99 &   141 &   133 &    31 &    29 \\ 
 & & $\Delta_{\rm NLTE}$ & $-$0.07 & $-$0.14 & $-$0.15 & $-$0.24 & $-$0.28 &  0.22 &  0.17 \\ 
 4750/2.00/$-$2.00 &  0.0 & 
 EW$_{\rm LTE}$ &   178 &   173 &    71 &   110 &   104 &    13 &    12 \\ 
 & & $\Delta_{\rm NLTE}$ & $-$0.09 & $-$0.18 & $-$0.06 & $-$0.18 & $-$0.21 &  0.18 &  0.16 \\ 
 5000/2.00/$-$2.00 &  0.5 & 
 EW$_{\rm LTE}$ &   202 &   191 &    91 &   128 &   122 &    30 &    28 \\ 
 & & $\Delta_{\rm NLTE}$ & $-$0.12 & $-$0.25 & $-$0.16 & $-$0.31 & $-$0.35 &  0.24 &  0.19 \\ 
 5000/2.00/$-$2.00 &  0.0 & 
 EW$_{\rm LTE}$ &   162 &   159 &    64 &   102 &    95 &    12 &    11 \\ 
 & & $\Delta_{\rm NLTE}$ & $-$0.14 & $-$0.24 & $-$0.04 & $-$0.19 & $-$0.22 &  0.20 &  0.18 \\ 
 5000/2.50/$-$2.00 &  0.5 & 
 EW$_{\rm LTE}$ &   196 &   184 &    82 &   121 &   114 &    23 &    21 \\ 
 & & $\Delta_{\rm NLTE}$ & $-$0.12 & $-$0.23 & $-$0.12 & $-$0.27 & $-$0.30 &  0.20 &  0.17 \\ 
 5000/2.50/$-$2.00 &  0.0 & 
 EW$_{\rm LTE}$ &   155 &   151 &    54 &    94 &    86 &     9 &     8 \\ 
 & & $\Delta_{\rm NLTE}$ & $-$0.13 & $-$0.20 & $-$0.02 & $-$0.15 & $-$0.17 &  0.18 &  0.16 \\ 
 4800/1.50/$-$3.00 &  1.1 & 
 EW$_{\rm LTE}$ &   199 &   189 &    90 &   127 &   121 &    24 &    23 \\ 
 & & $\Delta_{\rm NLTE}$ & $-$0.13 & $-$0.26 & $-$0.14 & $-$0.30 & $-$0.35 &  0.29 &  0.24 \\ 
 4800/1.50/$-$3.00 &  0.7 & 
 EW$_{\rm LTE}$ &   167 &   163 &    69 &   106 &   100 &    11 &    11 \\ 
 & & $\Delta_{\rm NLTE}$ & $-$0.14 & $-$0.26 & $-$0.03 & $-$0.19 & $-$0.23 &  0.25 &  0.23 \\ 
 5010/4.80/$-$3.40 &  1.1 & 
 EW$_{\rm LTE}$ &   100 &    82 &     8 &    32 &    25 &       &       \\ 
 & & $\Delta_{\rm NLTE}$ & $-$0.01 & $-$0.01 &  0.01 &  0.01 &  0.01 &       &       \\ 
 5050/2.34/$-$2.84 &  1.0 & 
 EW$_{\rm LTE}$ &   169 &   161 &    67 &   103 &    96 &    14 &    13 \\ 
 & & $\Delta_{\rm NLTE}$ & $-$0.17 & $-$0.30 & $-$0.08 & $-$0.28 & $-$0.30 &  0.24 &  0.21 \\ 
 5050/2.34/$-$2.84 &  0.5 & 
 EW$_{\rm LTE}$ &   136 &   132 &    39 &    78 &    71 &     5 &     5 \\ 
 & & $\Delta_{\rm NLTE}$ & $-$0.15 & $-$0.20 &  0.03 & $-$0.09 & $-$0.10 &  0.24 &  0.21 \\ 
 5260/2.75/$-$2.85 &  0.6 & 
 EW$_{\rm LTE}$ &   126 &   121 &    30 &    70 &    61 &       &       \\ 
 & & $\Delta_{\rm NLTE}$ & $-$0.13 & $-$0.15 &  0.04 & $-$0.04 & $-$0.05 &       &       \\ 
 5260/2.75/$-$2.85 &  0.3 & 
 EW$_{\rm LTE}$ &   111 &   104 &    17 &    53 &    44 &       &       \\ 
 & & $\Delta_{\rm NLTE}$ & $-$0.04 & $-$0.03 &  0.08 &  0.05 &  0.04 &       &       \\ 
\noalign{\smallskip} 
\end{longtable}
}

\longtab{8}{
\begin{longtable}[]{lcrrrrrrrrrrr}   
 \caption{\label{Tab:eu_nlte} The same as in Table\,\ref{Tab:ba_nlte} for lines of \ion{Eu}{ii}. Everywhere, \kH\ = 0.1.} \\
\hline\hline \noalign{\smallskip} 
Model & [Eu/Fe] & & \multicolumn{10}{c}{\ion{Eu}{ii} lines, $\lambda$(\AA)} \\
\cline{4-13} \noalign{\smallskip} 
   &  &         &  4129 &  4205 &  3819 &  3724 &  3907 &  4435 &  4522 &  3971 &  3930 &  6645 \\ 
\noalign{\smallskip} \hline \noalign{\smallskip} 
\endfirsthead
\caption{continued.}\\
\hline \noalign{\smallskip} 
Model & [Eu/Fe] & & \multicolumn{10}{c}{\ion{Eu}{ii} lines, $\lambda$(\AA)} \\
\cline{4-13} \noalign{\smallskip} 
   &  &         &  4129 &  4205 &  3819 &  3724 &  3907 &  4435 &  4522 &  3971 &  3930 &  6645 \\ 
\hline  \\   
\endhead
\hline
\endfoot
\hline
\endlastfoot
 4500/1.00/$-$3.00 &  1.5 & 
 EW$_{\rm LTE}$ &   183 &   132 &   176 &    81 &   128 &   100 &    39 &    75 &    74 &    16 \\ 
 & & $\Delta_{\rm NLTE}$ &  0.08 &  0.06 &  0.06 &  0.07 &  0.18 &  0.08 &  0.05 &  0.09 &  0.09 &  0.10 \\ 
 4500/1.00/$-$3.00 &  0.7 & 
 EW$_{\rm LTE}$ &    56 &    39 &    79 &    21 &    43 &    24 &     7 &    42 &    41 &     \\ 
 & & $\Delta_{\rm NLTE}$ &  0.17 &  0.13 &  0.16 &  0.16 &  0.31 &  0.18 &  0.14 &  0.27 &  0.27 &   \\ 
 4500/1.50/$-$3.00 &  1.5 & 
 EW$_{\rm LTE}$ &   158 &   113 &   166 &    65 &   114 &    80 &    29 &    69 &    69 &    11 \\ 
 & & $\Delta_{\rm NLTE}$ &  0.07 &  0.05 &  0.06 &  0.07 &  0.14 &  0.07 &  0.05 &  0.10 &  0.10 &  0.10 \\ 
 4500/1.50/$-$3.00 &  0.7 & 
 EW$_{\rm LTE}$ &    40 &    28 &    62 &    14 &    32 &    17 &     5 &    34 &    33 &     \\ 
 & & $\Delta_{\rm NLTE}$ &  0.15 &  0.12 &  0.14 &  0.15 &  0.24 &  0.16 &  0.13 &  0.24 &  0.23 &   \\ 
 4500/2.00/$-$3.00 &  1.5 & 
 EW$_{\rm LTE}$ &   129 &    91 &   152 &    49 &    95 &    60 &    20 &    63 &    63 &     8 \\ 
 & & $\Delta_{\rm NLTE}$ &  0.06 &  0.05 &  0.05 &  0.06 &  0.11 &  0.07 &  0.05 &  0.11 &  0.11 &  0.11 \\ 
 4500/2.00/$-$3.00 &  0.7 & 
 EW$_{\rm LTE}$ &    28 &    19 &    46 &     9 &    22 &    11 &     &    26 &    26 &     \\ 
 & & $\Delta_{\rm NLTE}$ &  0.13 &  0.11 &  0.12 &  0.13 &  0.19 &  0.15 &   &  0.19 &  0.19 &   \\ 
 4750/1.00/$-$3.00 &  1.5 & 
 EW$_{\rm LTE}$ &   150 &   107 &   161 &    61 &   110 &    76 &    27 &    68 &    68 &    12 \\ 
 & & $\Delta_{\rm NLTE}$ &  0.09 &  0.07 &  0.08 &  0.09 &  0.21 &  0.09 &  0.07 &  0.15 &  0.15 &  0.12 \\ 
 4750/1.00/$-$3.00 &  0.7 & 
 EW$_{\rm LTE}$ &    37 &    26 &    58 &    13 &    30 &    16 &     5 &    33 &    32 &     \\ 
 & & $\Delta_{\rm NLTE}$ &  0.18 &  0.15 &  0.18 &  0.19 &  0.28 &  0.18 &  0.15 &  0.28 &  0.28 &   \\ 
 4750/1.50/$-$3.00 &  1.5 & 
 EW$_{\rm LTE}$ &   122 &    86 &   145 &    46 &    92 &    58 &    19 &    62 &    62 &     9 \\ 
 & & $\Delta_{\rm NLTE}$ &  0.08 &  0.06 &  0.07 &  0.09 &  0.17 &  0.09 &  0.07 &  0.16 &  0.16 &  0.14 \\ 
 4750/1.50/$-$3.00 &  0.7 & 
 EW$_{\rm LTE}$ &    26 &    18 &    44 &     9 &    21 &    11 &     &    25 &    25 &     \\ 
 & & $\Delta_{\rm NLTE}$ &  0.15 &  0.13 &  0.16 &  0.17 &  0.23 &  0.17 &   &  0.24 &  0.24 &   \\ 
 4750/2.00/$-$3.00 &  1.5 & 
 EW$_{\rm LTE}$ &    95 &    67 &   126 &    34 &    74 &    43 &    14 &    55 &    55 &     6 \\ 
 & & $\Delta_{\rm NLTE}$ &  0.08 &  0.06 &  0.06 &  0.08 &  0.14 &  0.09 &  0.07 &  0.16 &  0.16 &  0.16 \\ 
 4750/2.00/$-$3.00 &  0.7 & 
 EW$_{\rm LTE}$ &    18 &    13 &    32 &     6 &    15 &     8 &     &    19 &    19 &     \\ 
 & & $\Delta_{\rm NLTE}$ &  0.13 &  0.12 &  0.14 &  0.15 &  0.19 &  0.16 &   &  0.20 &  0.20 &   \\ 
 5000/2.00/$-$3.00 &  1.5 & 
 EW$_{\rm LTE}$ &    66 &    46 &    98 &    23 &    54 &    29 &     9 &    47 &    46 &     5 \\ 
 & & $\Delta_{\rm NLTE}$ &  0.09 &  0.08 &  0.09 &  0.11 &  0.16 &  0.11 &  0.09 &  0.19 &  0.19 &  0.23 \\ 
 5000/2.00/$-$3.00 &  0.7 & 
 EW$_{\rm LTE}$ &    12 &     8 &    21 &     &    10 &     5 &     &    14 &    13 &     \\ 
 & & $\Delta_{\rm NLTE}$ &  0.14 &  0.13 &  0.16 &   &  0.19 &  0.16 &   &  0.20 &  0.20 &   \\ 
 5000/2.50/$-$3.00 &  1.5 & 
 EW$_{\rm LTE}$ &    47 &    33 &    76 &    16 &    39 &    21 &     6 &    39 &    38 &     \\ 
 & & $\Delta_{\rm NLTE}$ &  0.09 &  0.08 &  0.08 &  0.10 &  0.13 &  0.10 &  0.09 &  0.17 &  0.17 &   \\ 
 5000/2.50/$-$3.00 &  0.7 & 
 EW$_{\rm LTE}$ &     8 &     6 &    15 &     &     7 &     &     &    10 &    10 &     \\ 
 & & $\Delta_{\rm NLTE}$ &  0.12 &  0.12 &  0.14 &   &  0.16 &   &   &  0.17 &  0.17 &   \\ 
 4500/1.00/$-$2.00 &  0.7 & 
 EW$_{\rm LTE}$ &   193 &   139 &   186 &    85 &   136 &   107 &    43 &    78 &    77 &    17 \\ 
 & & $\Delta_{\rm NLTE}$ &  0.06 &  0.04 &  0.05 &  0.04 &  0.14 &  0.06 &  0.03 &  0.05 &  0.05 &  0.07 \\ 
 4500/1.50/$-$2.00 &  0.7 & 
 EW$_{\rm LTE}$ &   161 &   115 &   169 &    65 &   116 &    81 &    29 &    71 &    70 &    12 \\ 
 & & $\Delta_{\rm NLTE}$ &  0.06 &  0.04 &  0.05 &  0.04 &  0.12 &  0.06 &  0.03 &  0.06 &  0.06 &  0.08 \\ 
 4500/2.00/$-$2.00 &  0.7 & 
 EW$_{\rm LTE}$ &   125 &    88 &   148 &    47 &    93 &    58 &    19 &    63 &    63 &     8 \\ 
 & & $\Delta_{\rm NLTE}$ &  0.06 &  0.04 &  0.04 &  0.04 &  0.10 &  0.06 &  0.04 &  0.08 &  0.08 &  0.09 \\ 
 4750/1.00/$-$2.00 &  0.7 & 
 EW$_{\rm LTE}$ &   176 &   126 &   181 &    73 &   128 &    93 &    34 &    73 &    73 &    16 \\ 
 & & $\Delta_{\rm NLTE}$ &  0.07 &  0.04 &  0.06 &  0.05 &  0.17 &  0.07 &  0.03 &  0.07 &  0.07 &  0.09 \\ 
 4750/1.50/$-$2.00 &  0.7 & 
 EW$_{\rm LTE}$ &   141 &   100 &   161 &    55 &   106 &    69 &    24 &    67 &    66 &    11 \\ 
 & & $\Delta_{\rm NLTE}$ &  0.07 &  0.04 &  0.06 &  0.05 &  0.14 &  0.07 &  0.04 &  0.08 &  0.08 &  0.10 \\ 
 4750/2.00/$-$2.00 &  0.7 & 
 EW$_{\rm LTE}$ &   106 &    75 &   137 &    38 &    82 &    49 &    16 &    59 &    59 &     7 \\ 
 & & $\Delta_{\rm NLTE}$ &  0.07 &  0.04 &  0.05 &  0.05 &  0.12 &  0.07 &  0.05 &  0.10 &  0.10 &  0.12 \\ 
 5000/2.00/$-$2.00 &  0.7 & 
 EW$_{\rm LTE}$ &    84 &    59 &   119 &    29 &    68 &    38 &    12 &    53 &    53 &     6 \\ 
 & & $\Delta_{\rm NLTE}$ &  0.08 &  0.05 &  0.07 &  0.07 &  0.14 &  0.09 &  0.06 &  0.13 &  0.13 &  0.16 \\ 
 5000/2.50/$-$2.00 &  0.7 & 
 EW$_{\rm LTE}$ &    59 &    41 &    92 &    20 &    49 &    26 &     8 &    44 &    44 &     \\ 
 & & $\Delta_{\rm NLTE}$ &  0.08 &  0.06 &  0.06 &  0.07 &  0.12 &  0.09 &  0.06 &  0.12 &  0.12 &   \\ 
 4800/1.50/$-$3.00 &  1.5 & 
 EW$_{\rm LTE}$ &   117 &    83 &   142 &    44 &    89 &    56 &    18 &    61 &    61 &     9 \\ 
 & & $\Delta_{\rm NLTE}$ &  0.09 &  0.07 &  0.08 &  0.09 &  0.18 &  0.09 &  0.07 &  0.17 &  0.17 &  0.15 \\ 
 5010/4.80/$-$3.40 &  1.9 & 
 EW$_{\rm LTE}$ &     6 &       &    11 &       &     5 &       &       &     7 &     7 &       \\ 
 & & $\Delta_{\rm NLTE}$ &  0.03 &       &  0.03 &       &  0.04 &       &       &  0.05 &  0.05 &       \\ 
 5050/2.34/$-$2.84 &  1.0 & 
 EW$_{\rm LTE}$ &    23 &    16 &    40 &     8 &    20 &    10 &       &    24 &    23 &       \\ 
 & & $\Delta_{\rm NLTE}$ &  0.13 &  0.12 &  0.14 &  0.14 &  0.18 &  0.14 &       &  0.18 &  0.18 &       \\ 
 5260/2.75/$-$2.85 &  1.1 & 
 EW$_{\rm LTE}$ &    14 &    10 &    26 &     5 &    12 &     6 &       &    17 &    16 &       \\ 
 & & $\Delta_{\rm NLTE}$ &  0.12 &  0.11 &  0.12 &  0.13 &  0.15 &  0.13 &       &  0.17 &  0.17 &       \\ 
\noalign{\smallskip} 
\end{longtable}
}

\end{document}